# A Unified Theory and Fundamental Rules of Strain-dependent Doping Behaviors in Semiconductors

Xiaolan Yan[1], Pei Li[1], Su-Huai Wei[1,2*], and Bing Huang[1,2*]

[1] *Beijing Computational Science Research Center, Beijing 100193, China*
[2] *Department of Physics, Beijing Normal University, Beijing 100875, China*

E-mails: suhuaiwei@csrc.ac.cn; bing.huang@csrc.ac.cn

**Enhancing the dopability of semiconductors via strain engineering is critical to improving their functionalities, which is, however, largely hindered by the lack of fundamental rules. In this Letter, for the first time, we develop a unified theory to understand the total energy changes of defects (or dopants) with different charge states under strains, which can exhibit either parabolic or superlinear behaviors, determined by the size of defect-induced local volume change ($\Delta V$). In general, $\Delta V$ increases (decreases) when an electron is added (removed) to (from) the defect site. Consequently, in terms of this unified theory, three fundamental rules can be obtained to further understand or predict the diverse strain-dependent doping behaviors, *i.e.*, defect formation energies, charge-state transition levels, and Fermi pinning levels, in semiconductors. These three fundamental rules could be generally applied to improve the doping performance or overcome the doping bottlenecks in various semiconductors.**

The application of semiconductors in electronic and optoelectronic devices critically depends on their dopability. Generally, there are three important factors that can fundamentally limit the dopability in semiconductors: (i) the desirable defects or dopants (generally denoted as defects hereafter) have limited solubility, *i.e.*, their formation energies ($H_f^{D,q}$) are too high [1,2]; (ii) the desirable defects have sufficient solubility, but they are too difficult to be ionized at room temperature, *i.e.*, their charge-state transition levels ($\varepsilon^{q/q'}$) are too deep inside the bandgap [1,2]; and (iii) the desirable defects have low ($H_f^{D,q}$) and shallow $\varepsilon^{q/q'}$, unfortunately, the intrinsic compensating defects can easily form and pin the Fermi-level position ($E_{pin}$) deep inside the bandgap, preventing the further increase of desired free carriers [1,2]. Comparing to (i) and (ii), (iii) is the most difficult one to be overcome as it belongs to the intrinsic property of semiconductors.

In the past decades, strain engineering is widely adopted to enhance the performances of semiconductors, *e.g.*, optimize the electronic structures [3-7], improve phase stabilities [8,9], generate spin currents [10], control carrier excitations or transports [11-13], and modulate ion diffusion paths [14,15]. It is not surprising that strain engineering has also been used to tune the doping performances in semiconductors [16-27]. However, it is rather puzzled that the strain-induced changes of doping behaviors for different defects in different semiconductors are dramatically different [16,18-24,26]. Unfortunately, a unified theory that can intuitively understand all these diverse doping behaviors in different systems is still lacking, which prevents us to establish the fundamental rules to overcome the doping bottlenecks in semiconductors.

In this Letter, we have developed a simple but *unified* theory for understanding the strain-dependent total energy changes of defects ($\Delta E_t^{D,q}$) under different charge states, which is critically determined by the defect-induced local volume change ($\Delta V$). Depending on the size of $\Delta V$, the $\Delta E_t^{D,q}$ of a defect can exhibit either parabolic ($\Delta V \sim 0$) or monotonic ($dE_t^{D,q}/dV \sim -\Delta V$) dependences. Noticeably, the $\Delta V$ is *q*-dependent, which increases (decreases) for more negatively (positively) charged defects. Based on this unified theory of $\Delta E_t^{D,q}$, we can establish *three fundamental rules* on understanding the strain-dependent $H_f^{D,q}$, $\varepsilon^{q/q'}$, and $E_{pin}$, which may consequently be applied to overcome the above (i)-(iii) doping problems in various semiconductor systems.

***A unified theory on*** $\Delta E_t^{D,q}$***.*** For a system without a defect, its total energy $E_t^{host}(V)$ as a function of volume ($V$), to the lowest order, follows: $E_t^{host}(V)=\alpha_0(V-V_0)^2$, where $V_0$ is the equilibrium volume of host lattice and $\alpha_0=\frac{1}{2}B_0/V_0$, with $B_0$ being the bulk modulus. Similarly, for a system with a defect in the $q$ charge state, its total energy $E_t^{D,q}(V)$ follows: $E_t^{D,q}(V) = (\alpha_0+\Delta\alpha)[V-(V_0+\Delta V)]^2$, where $\Delta\alpha$ and $\Delta V$ are the changes of $\alpha_0$ and $V_0$ induced by the defect, respectively [20]. Ignoring the high order terms, the total energy changes $\Delta E_t^{D,q}$ induced by the defect as a function of $V$ can be derived as:

$$\Delta E_t^{D,q}(V) = E_t^{D,q}(V) - E_t^{host}(V) = -2\alpha_0\Delta V(V - V_0) + \Delta\alpha(V - V_0)^2. \tag{1}$$

Obviously, $\Delta E_t^{D,q}(V)$ is determined by the two terms that are mainly associated with $\Delta\alpha$ and $\Delta V$ in Eq. (1). The $\Delta\alpha$ is usually negligible, especially for substitutional defects where the chemical and size differences are small [28]. The $\Delta V$ depends on the size difference between the dopant and the host element, therefore, is noticeable in most cases [19,20,24]. In these cases, $\Delta E_t^{D,q}(V)$ is largely determined by the first term of Eq. (1), giving rise to a linear dependence on $V$ (**Fig. 1a**). However, if the defect-induced $\Delta V$ is not significant, the high-order second term in Eq. (1) could become dominant, giving rise to a parabolic change of $\Delta E_t^{D,q}(V)$ under strain (**Fig. 1b**) [21-23]. Importantly, since $\Delta V$ can be rather sensitive to the charge states of a defect, it is expected that dramatically different $q$-dependent behaviors of $\Delta E_t^{D,q}(V)$ could exist even for the same defect in a semiconductor.

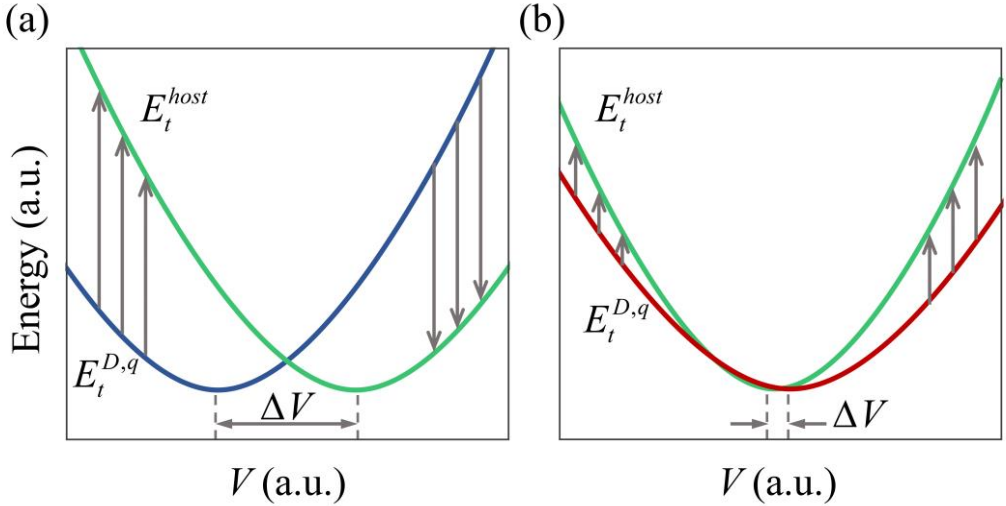

**Fig 1.** Schematic plotting of total energy as a function of volume $V$ for a host with $[E_t^{D,q}(V)]$ and without $[E_t^{host}(V)]$ a defect. $\Delta E_t^{D,q}(V)$ is indicated by the black arrows, which are mostly determined by the (a) first and (b) second terms of Eq. (1), respectively.

The defect formation energy $H_f^{D,q}(V, E_F)$ as a function of $V$ can be written as [2,29-31]

$$H_f^{D,q}(V, E_F) = \Delta E_t^{D,q}(V) + \sum n_i\mu_i + qE_F, \tag{2}$$

where $E_F$ is the absolute Fermi energy in the bandgap, $\mu_i$ is the chemical potential referenced to the total energy of elements in its lowest energy bulk form. Therefore, $H_f^{D,q}(V)$ follows the same trend as $\Delta E_t^{D,q}(V)$ as a function of $V$ at a given $E_F$. The $\varepsilon^{q/q'}(V)$ referenced to the valence band maximum (VBM) energy $E_{VBM}(V)$ is the Fermi energy at which the same defect $D$ with different charge state $q$ and $q'$ have the same formation energy, therefore, it can be described as

$$\varepsilon^{q/q'}(V) = \left\{\frac{\Delta E_t^{D,q}(V) - \Delta E_t^{D,q'}(V)}{q' - q} - E_{VBM}(V_0)\right\} - \Delta E_{VBM}(V). \tag{3}$$

As shown in Eq. (3), the $\varepsilon^{q/q'}(V)$ includes two terms: the first term represents the absolute value of $\varepsilon^{q/q'}(V)$ [denoted as $a$-$\varepsilon^{q/q'}(V)$] referenced to the VBM energy of the unstrained host, and the second term is the $V$-dependent shift of the VBM energy, $\Delta E_{VBM}(V)$. It should be noticed that in a semiconductor, the $\Delta E_{VBM}(V)$ depends on its absolute volume deformation potentials (AVDP). For most common semiconductors with a dominant bonding VBM state, the AVDP is positive [32], *i.e.*, the $\Delta E_{VBM}(V)$ will increase (decrease) under tensile (compressive) strains. Therefore, the understanding of $V$-dependent $a$-$\varepsilon^{q/q'}(V)$ is the key for the understanding of the trend of $\varepsilon^{q/q'}(V)$. Based on the above discussions, in the following, we will derive three fundamental rules of strain-dependent doping behaviors in semiconductors with numerical verifications.

***Rule No. I: strain-dependent $H_f^{D,q}$.*** Taking GaN as a typical example, we have employed the first-principles calculations (See computational methods in Supplemental Material (SM) [33]) to study the uniform strain $\eta$ on the change of $H_f^{D,q}$ [$\Delta H_f^{D,q}(\eta)$] for different defects in this system. Firstly, we consider N vacancy ($V_N$), the dominant intrinsic defect in GaN [34,35]. As shown in **Fig. 2a**, the $\Delta H_f^{D,q}(\eta)$ of $V_N$ in its neutral charge state ($V_N^0$) exhibits mostly a parabolic dependence of $\eta$, *i.e.*, its $H_f^{D,q}(\eta)$ tends to decrease under both compressive and tensile $\eta$. Therefore, the $\Delta H_f^{D,q}(\eta)$ of $V_N^0$ could be mostly determined by the second term of Eq. (1) (**Fig. 1b**). Indeed, our calculation confirms that a rather small value of $\Delta V$ ($\Delta V \sim -1.87$ Å$^3$/$V_N^0$) exists for $V_N^0$ (**Fig. 2d**). The negative sign of $\Delta \alpha \sim -0.06$ suggests that the local bulk modulus is reduced with the formation of $V_N$, as expected. Interestingly, when $V_N^0$ is converted to its +1 state ($V_N^{+1}$), its local volume is significantly reduced to $\Delta V \sim -7.3$ Å$^3$/$V_N^{+1}$ due to reduced charge occupation, giving rise to a large left-shift of its energy curve in **Fig. 2d**. Consequently, differing from $V_N^0$, the $\Delta H_f^{D,q}(\eta)$ of $V_N^{+1}$ (at a given $E_F$) is now mainly determined by the first term of Eq. (1) (**Fig. 1a**), *i.e.*, $H_f^{D,q}(\eta)$ linearly decreases (increases) under compressive (tensile) $\eta$. Compared to $V_N^{+1}$, the linear slope of $\Delta H_f^{D,q}(\eta)$ for $V_N^{+3}$ is only slightly larger than that for $V_N^{+1}$ due to the similar $\Delta V$ in both charge states, a reflection of Coulomb interaction between the defects and its local environment, as shown in **Fig. 2a**. The similar $q$-dependent $\Delta H_f^{D,q}(\eta)$ have been also observed for the intrinsic defects in other semiconductors, *e.g.*, $V_C$ in SiC (**Fig. S1a** [33]), $V_{Ga}$ in GaN (**Fig. S1b** [33]), and $V_{Zn}$ in ZnTe (**Fig. S2a** [33]).

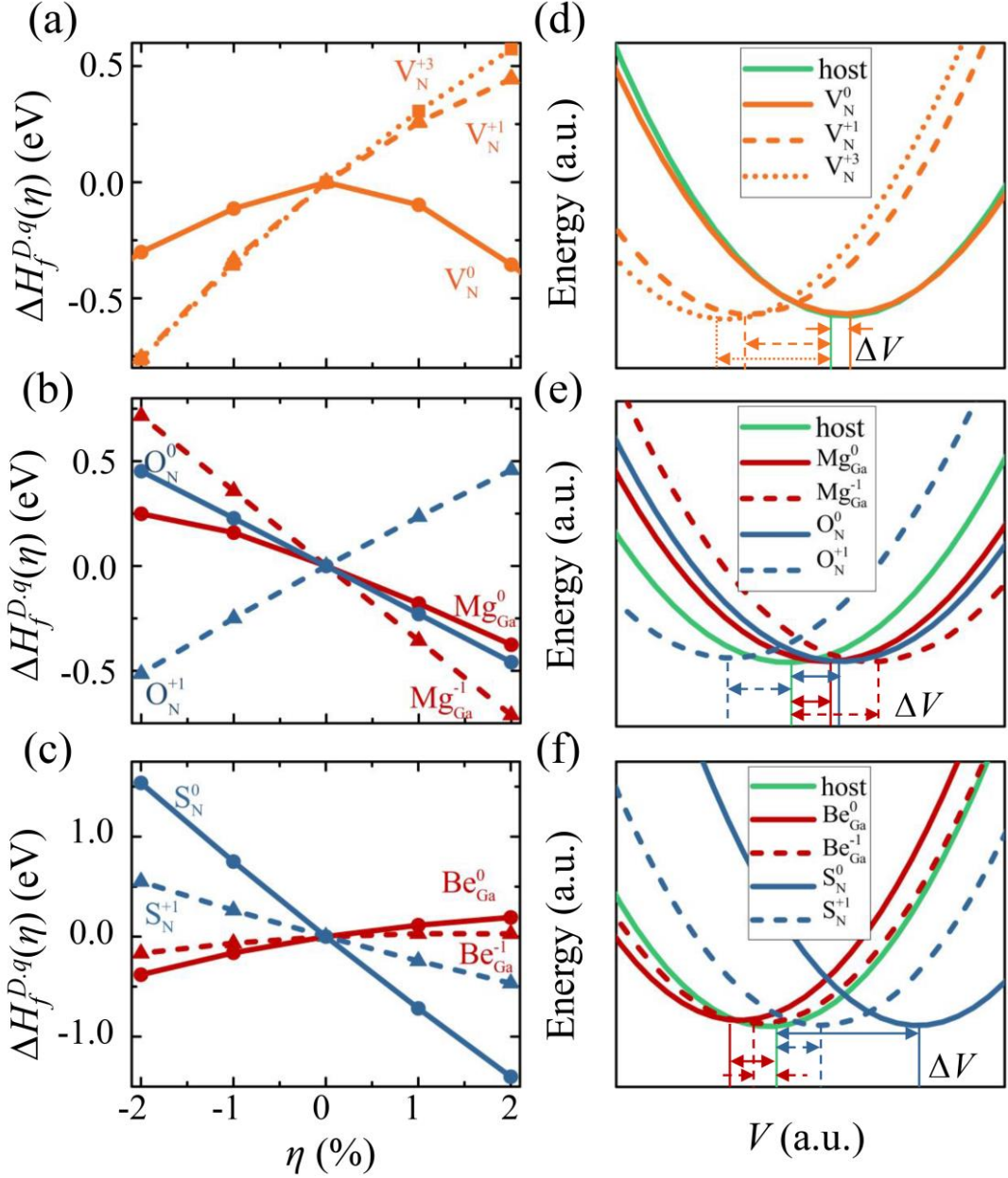

**Fig 2.** Rule No. I on strain-dependent $H_f^{D,q}$. (a) Change of formation energies $\Delta H_f^{D,q}(\eta)$ for N vacancy ($V_N$) in GaN as a function of strain $\eta$. (b) Similar to (a) but for *n*-type ($O_N$) and *p*-type ($Mg_{Ga}$) dopants in GaN. (c) Similar to (a) but for *n*-type ($S_N$) and *p*-type ($Be_{Ga}$) dopants in GaN. Schematic plotting of total energies $E_t^{D,q}(V)$ as a function of volume $V$ for (d) $V_N$, (e) $O_N$ and $Mg_{Ga}$, and (f) $S_N$ and $Be_{Ga}$, respectively. $E_t^{host}(V)$ for host are also shown in (d)-(f) for comparison.

Secondly, we consider the substitutional doping in GaN. In order to understand the size effects of dopants, O and S are selected as *n*-type dopants, while Mg and Be are selected as *p*-type ones. For the case of $O_N^0$, the electronic environment around the anion site induces a positive $\Delta V \sim +5.59$ Å$^3$/$O_N^0$ (**Fig. 2e**). As shown in **Fig. 2b**, the $\Delta H_f^{D,q}(\eta)$ of $O_N^0$ is dominated by the first term of Eq. (1), resulting in a linear increase (decrease) as a function of compressive (tensile) $\eta$. When $O_N^0$ is ionized to $O_N^{+1}$, its $\Delta V$ largely shrinks to a negative value of $\Delta V \sim -5.75$ Å$^3$/$O_N^{+1}$, which can largely left-shift its energy curve in **Fig. 2e**. Consequently, the linear slope of $\Delta H_f^{D,q}(\eta)$ for $O_N^{+1}$ is inverted compared to that of $O_N^0$. Differing from $O_N$, as shown in **Fig. 2f**, the larger ionic size of S than N induces a much larger $\Delta V$ after $S_N$ doping ($\Delta V \sim +17.72$ Å$^3$/$S_N^0$), giving rise to a much larger linear slope of $\Delta H_f^{D,q}(\eta)$, as shown in **Fig. 2c**. When $S_N^0$ is ionized to $S_N^{+1}$, its $\Delta V$ largely shrinks to a (still positive) value of $\sim +6.13$ Å$^3$/$S_N^{+1}$, with a significant left-shift of its energy curve in **Fig. 2f**. As a result, the linear slope of $S_N^{+1}$ is reduced compared to that of $S_N^0$. Therefore, depending on the initial different $\Delta V$ at neutral charge states, the *n*-type $O_N^{+1}$ and $S_N^{+1}$ can have an *opposite* linear dependence of $\Delta H_f^{D,q}(\eta)$. However, $\Delta V$ always decreases when electron is removed from the dopant site (more positively charged) and $H_f^{D,q}(\eta)$ follows $dE_t^{D,q}(\eta)/dV \sim -\Delta V$.

Thirdly, we consider the *p*-type doping in GaN. The $\Delta V$ of $Mg_{Ga}^0$ is positive ($\Delta V \sim +4.06$ Å$^3$/$Mg_{Ga}^0$) (**Fig. 2e**) due to the larger ionic size of Mg than Ga, which can further expand when it is ionized to $Mg_{Ga}^{-1}$ ($\Delta V \sim +8.73$ Å$^3$/$Mg_{Ga}^{-1}$). Therefore, as shown in **Fig. 2b**, the $Mg_{Ga}^{-1}$ can have a similar linear but enlarged slope effect after it is ionized. On the other hand, $\Delta V$ for $Be_{Ga}^0$ ($\Delta V \sim -3.26$ Å$^3$/$Be_{Ga}^0$) is negative due to the small size of Be and after it is negatively charged, forming $Be_{Ga}^{-1}$, again, the $\Delta V$ increases to $\sim -1.07$ Å$^3$/$Be_{Ga}^{-1}$. Because of the rather small (absolute) value of $\Delta V$, $\Delta H_f^{D,q}(\eta)$ of $Be_{Ga}^{-1}$ exhibits mostly a parabolic dependence of $\eta$, as shown in **Fig. 2c**, dramatically differing from that of $Be_{Ga}^0$. Therefore, depending on the initial different $\Delta V$ at neutral charge states, the *p*-type $Mg_{Ga}^{-1}$ and $Be_{Ga}^{-1}$ can have either linear or parabolic dependence of $\Delta H_f^{D,q}(\eta)$. In general, the diverse trends of *q*-dependent $\Delta H_f^{D,q}(\eta)$ for different dopants in different semiconductors, *e.g.*, $Cl_{Te}$ in ZnTe (**Fig. S2b** [33]), $Al_{Si}$ and $N_C$ in SiC (**Fig. S3** [33]), $C_N$ and $Ge_{Ga}$ in GaN (**Fig. S4a-b** [33]), $Zn_{Ga}$, $Si_P$, $Ge_{Ga}$, and $S_P$ in GaP (**Fig. S4c-d** [33]) and $As_{Si}$ in Si [19], could be understood in a similar way via tracking the evolution of $\Delta V$.

From the analysis above, we can reach the Rule No. I on *η*-dependent $\Delta H_f^{D,q}(\eta)$ in semiconductors: *$\Delta H_f^{D,q}(\eta)$ of defects under different charge states can have either parabolic ($\Delta V\sim 0$) or linear ($\Delta V \neq 0$) dependence, i.e., $dE_t^{D,q}(V)/dV \sim -\Delta V$. Here, the defect-induced $\Delta V$ is q-dependent, it increases (decreases) when electron is added (removed) to (from) the dopant site.* We want to emphasize that this rule is independent of the sizes of supercell calculations (**Fig. S5** [33]).

***Rule No. II: strain-dependent $\varepsilon^{q/q'}$.*** According to Eq. (3), the $\eta$-dependent $a$-$\varepsilon^{q/q'}(\eta)$, i.e., without consideration of $\Delta E_{VBM}(V)$, is also determined by $\Delta E_t^{D,q}(\eta)$. In general, for a *p*-type acceptor (**Fig. 3a**), in order to increase (decrease) the value of $a$-$\varepsilon^{0/-1}$, one needs to increase (decrease) its $H_f^{D,-1}$ more than $H_f^{D,0}$, which can be achieved under a negative (positive) $\eta$ in terms of the *Rule No. I*. Similarly, for a *n*-type donor (**Fig. 3a**), in order to increase (decrease) the value of $a$-$\varepsilon^{0/+1}$, one needs to increase (decrease) its $H_f^{D,0}$ more than $H_f^{D,+1}$, which can also be achieved under a negative (positive) $\eta$. As shown in **Fig. 3b**, the calculated $a$-$\varepsilon^{0/-1}(\eta)$ of Mg$_{Ga}$ and $a$-$\varepsilon^{0/+1}(\eta)$ of O$_N$ in GaN confirm our analysis, *i.e.*, the $a$-$\varepsilon^{0/-1}$ ($a$-$\varepsilon^{0/+1}$) moves towards VBM (CBM) under a tensile (compressive) $\eta$. We emphasize that this trend is general for all the defects due to the generality of *Rule No. I*. It is also independent of the $\Delta V$ of defects at their neutral charge states, *e.g.*, see the cases of Be$_{Ga}$ and S$_N$ in GaN (**Fig. S6** [33]) and Al$_{Si}$ and N$_C$ in SiC (**Fig. S7** [33]). The above discussions can drive us to the *Rule No. II* that *the negative (positive) $\eta$ is always beneficial for the realization of shallower $a$-$\varepsilon^{q/q'}(\eta)$ for the donor (acceptor) in semiconductors.* As discussed later, it should be noticed that the shallowness of $\varepsilon^{q/q'}(\eta)$ relative to the VBM of the strained system depends also on the shift $\Delta E_{VBM}(V)$ as shown in Eq. (3).

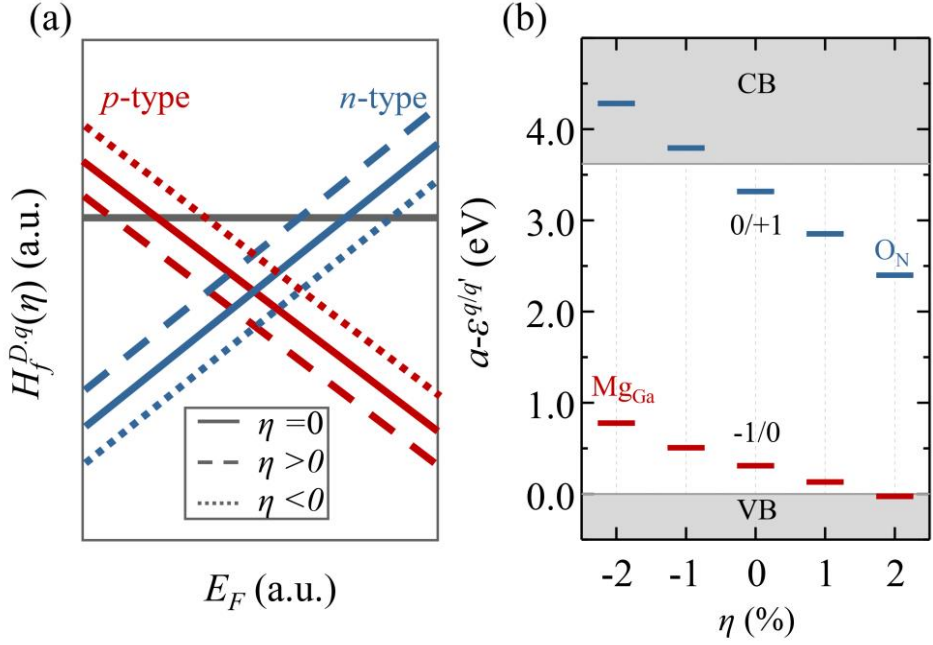

**Fig. 3** Rule No. II on strain-dependent $a$-$\varepsilon^{q/q'}$. (a) Schematic plotting of the change of formation energy $H_f^{D,q}(\eta)$ for positively-charged donor (blue) and negatively-charged acceptor (red), with respect to their neutral states (gray), under different $\eta$ as a function of $E_F$. (b) $a$-$\varepsilon^{0/-1}(\eta)$ and $a$-$\varepsilon^{0/+1}(\eta)$ for $Mg_{Ga}$ and $O_N$ in GaN as a function of strain $\eta$, respectively. Here the band edge positions are fixed at the values of $E_{VBM}(V_0)$ and $E_{CBM}(V_0)$.

***Rule No. III: strain-dependent $E_{pin}$.*** The *Rule Nos. I-II* demonstrate that the strain can always induce opposite changes of $H_f^{D,q}(\eta)$ for donors and acceptors, which can be utilized to tuning the Fermi level pinning $E_{pin}$ positions. The $E_{pin}$ level is determined by the Fermi energy at which the compensating donor and acceptor defects holding opposite charge states have the same energy. It is known that *n*-type $E_{pin-n}$ and *p*-type $E_{pin-p}$ positions set up the doping limits of *n*-type and *p*-type doping, respectively, in a semiconductor, which is an intrinsic problem of semiconductors that are difficult to overcome [31].

Basically, based on the doping limit rules, a semiconductor with low VBM, *e.g.* GaN [31], is difficult to be doped *p*-type, while a semiconductor with high CBM, *e.g.* ZnTe [36], is difficult to be doped *n*-type. For GaN, Mg ($Mg_{Ga}$) has been widely selected as a *p*-type dopant [2,34,37], which has an $\varepsilon^{0/-1}$ at VBM+0.3 eV (**Fig. 4b**). However, when Fermi level $E_F$ is shifted towards VBM after $Mg_{Ga}$ doping, the spontaneous formation of $V_N^{+3}$ can compensate the *p*-type doping induced by $Mg_{Ga}^{-1}$, giving rise to a deep $E_{pin-p}$ position locating at VBM+0.7 eV, agreeing with previous calculations [35]. As shown in **Fig. 4a**, although the $\varepsilon^{0/-1}$ of $Mg_{Ga}$ is relatively shallow, its *p*-type doping performance is strongly downgraded by the formation of $V_N^{+3}$. To reduce $E_{pin-p}$ position in GaN, one needs to increase the $H_f^{D,q}$ of $V_N^{+3}$ but decrease the $H_f^{D,q}$ of $Mg_{Ga}^{-1}$, which can be achieved under a tensile-$\eta$-induced synergistic effect. Indeed, as shown in **Fig. 4c**, our calculations confirm that the $H_f^q(\eta)$ of $V_N^{+3}$ ($Mg_{Ga}^{-1}$) can gradually increase (decrease) as a function of tensile $\eta$, linearly shifting the absolute $E_{pin-p}(\eta)$ [$a$-$E_{pin-p}(\eta)$, without consideration of $\Delta E_{VBM}(V)$] towards lower energy positions.

For the case of ZnTe, Cl ($Cl_{Te}$) is commonly used as a *n*-type dopant [38,39] with a calculated $\varepsilon^{0/+1}$ at CBM-0.43 eV, agreeing with the previous calculations [40]. Unfortunately, as shown in **Fig. 4b**, the spontaneous formation of $V_{Zn}^{-2}$ can largely pin the $E_{pin-n}$ position deeply inside the bandgap, *i.e.*, $E_{pin-n}$=CBM-0.84 eV, preventing the ionization of $Cl_{Te}$. Similarly, as shown in **Fig. 4d**, a compressive-$\eta$-induced synergistic effect may decrease the $H_f^{D,q}$ of $Cl_{Te}^{+1}$ but increase the $H_f^{D,q}$ of $V_{Zn}^{-2}$, giving rise to the shift of $a$-$E_{pin-n}$ towards higher energy positions. Based on the above understandings, we can arrive at the *Rule No. III* on the $\eta$-dependent $a$-$E_{pin}(\eta)$ in semiconductors: *the tensile (compressive) strain is always beneficial for the realization of shallower $a$-$E_{pin-p}(\eta)$ [$a$-$E_{pin-n}(\eta)$] in semiconductors*. Again, as discussed later, the shallowness of $E_{pin-p}(\eta)$ [$E_{pin-n}(\eta)$] relative to the VBM (CBM) of the strained system depends also on the shift $\Delta E_{VBM}(V)$ [$\Delta E_{CBM}(V)$]. In addition, We emphasize that the *Rule Nos. I-III* are generally valid for semiconductors under the biaxial strain (**Fig. S8** [33]), because Rule No. I depends solely on the defect-induced $\Delta V$.

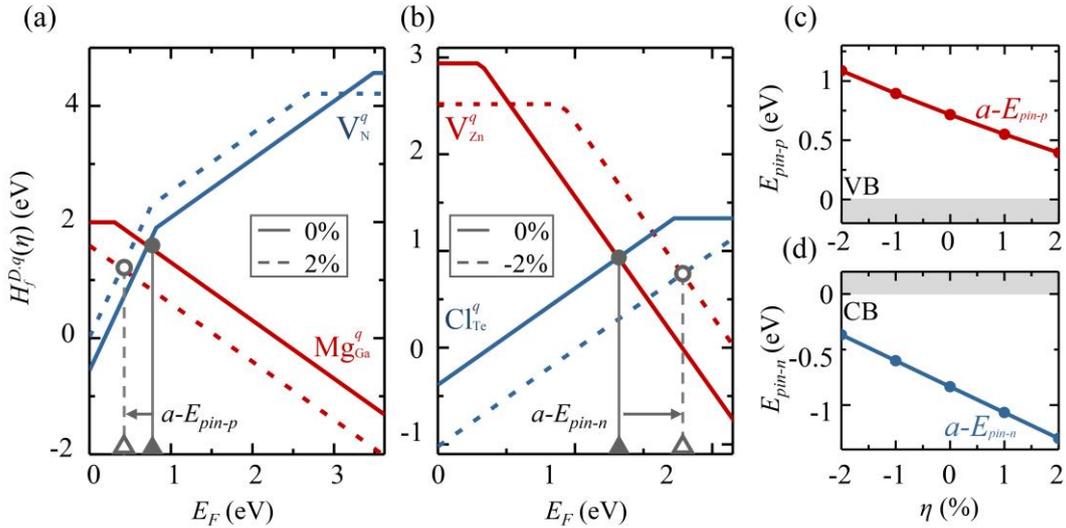

**Fig. 4** Rule No. III on strain-dependent $a$-$E_{pin}$. (a) Formation energy $H_f^{D,q}$ of external $Mg_{Ga}$ and intrinsic compensating $V_N$ in GaN without and with a +2% strain. (b) Formation energy $H_f^{p,q}$ of external $Cl_{Te}$ and intrinsic compensating $V_{Zn}$ in ZnTe without and with a -2% strain. $a$-$E_{pin}$ positions are marked by the vertical lines in (a)-(b). (c) $a$-$E_{pin\text{-}p}(\eta)$ and (d) $a$-$E_{pin\text{-}n}(\eta)$ positions as a function of $\eta$ in GaN and ZnTe, respectively. Here the band edge positions are fixed at the values of $E_{VBM}(V_0)$ and $E_{CBM}(V_0)$.

**Role of $\Delta E_{VBM}(V)/\Delta E_{CBM}(V)$.** Since the *Rule Nos. II-III* are for $a$-$\varepsilon^{q/q'}(\eta)$ and $a$-$E_{pin}(\eta)$, in order to obtain the rules of $\varepsilon^{q/q'}(\eta)$ and $E_{pin}(\eta)$ with respect to the band edge states under strain, the $\Delta E_{VBM}(V)/\Delta E_{CBM}(V)$ needs to be taken into account. As shown in **Fig. 5a**, in most common semiconductors with VBM (CBM) as bonding (antibonding) states, the $\Delta E_{VBM}(V)$ [$\Delta E_{CBM}(V)$] will almost linearly increase (decrease) as a function of $V$ from compression to tension [32]. Therefore, combining both *Rule Nos. II-III* and **Fig. 5a**, we can easily reach the conclusion that the *Rule Nos. II-III* also applies for the relative $\varepsilon^{q/q'}(\eta)$ of acceptors and $E_{pin\text{-}p}(\eta)$, because the opposite trend of $\eta$-dependent $a$-$\varepsilon^{0/-1}(\eta)$ [$a$-$E_{pin\text{-}p}(\eta)$] and $\Delta E_{VBM}(V)$ can induce a novel synergistic effect to make the $\varepsilon^{q/q'}(\eta)$ [$E_{pin\text{-}p}(\eta)$] of acceptors even shallower under a tensile strain. Indeed, as shown in **Figs. 5b-5c**, our calculations on $\varepsilon^{q/q'}$ and $E_{pin\text{-}p}$ in GaN confirm our analysis. The calculated $\varepsilon^{0/-1}$ [$E_{pin\text{-}p}$] of $Mg_{Ga}$ is shifted from VBM+0.3 [VBM+0.7] eV to VBM-0.15 [VBM+0.27] eV under $\eta$=+2%, significantly shallower than that of $a$-$\varepsilon^{0/-1}(\eta)$ [$a$-$E_{pin\text{-}p}$]. The trend of $\varepsilon^{q/q'}$ shown in **Fig. 5b** is consistent with the experimental observations [41]. We emphasize that there are no similar rules for $\varepsilon^{q/q'}(\eta)$ of donors and $E_{pin\text{-}n}(\eta)$ in common semiconductors, because of the $\eta$-dependent $a$-$\varepsilon^{0/+1}(\eta)$ [$a$-$E_{pin\text{-}n}(\eta)$] and $\Delta E_{CBM}(V)$ shift in the same direction, so the relative shift will depend on their individual changes.

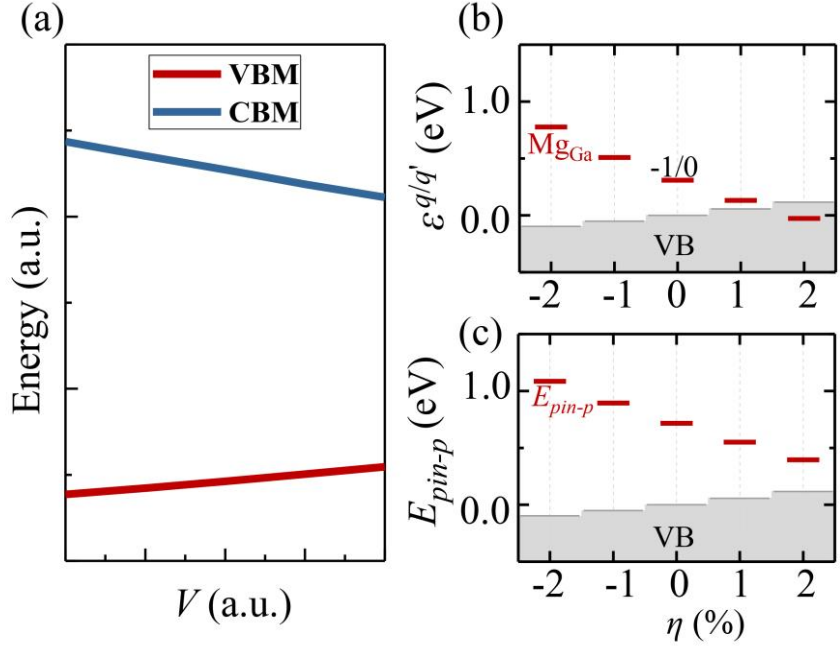

**Fig. 5** Role of $\Delta E_{VBM}(V)/\Delta E_{CBM}(V)$. (a) Schematic plotting of the band edge changes as a function of $V$ for common semiconductors with VBM (CBM) as bonding (antibonding) states. (b) $\varepsilon^{0/-1}(\eta)$ of $Mg_{Ga}$ and (c) $E_{pin-p}(\eta)$ in GaN as a function of strain $\eta$, with consideration of $\Delta E_{VBM}(V)$.

In summary, we have developed a unified theory and consequently established three fundamental rules for understanding the diverse strain-dependent doping behaviors in semiconductors, which can be applied to tune their $H_f^{p,q}$, $\varepsilon^{q/q'}$, and $E_{pin}$, as successfully confirmed by the first-principles calculations on several exemplary semiconductors. Generally, these fundamental rules can be widely applied to control doping and simultaneously overcome the doping bottlenecks in semiconductors via simple strain engineering.

**Acknowledgements:** The authors thank Drs. L. Kang, L. Hu, and J. F. Wang for helpful discussions. We acknowledge the support from and the NSFC under Grant Nos. 11634003, 11991060, and U1930402. S.H.W. also thank the support of the Key Research & Development Program of Beijing (Grant No: Z181100005118003). The calculations were performed at Tianhe2-JK at CSRC.